\documentclass[aps]{revtex4-1}
\usepackage{graphicx}

\usepackage{natbib}

\newcommand{\be}{\begin{equation}}
\newcommand{\ee}{\end{equation}}
\newcommand{\nn}{\mbox{} \nonumber \\ \mbox{} }
\newcommand{\ba}{\begin{eqnarray}}
\newcommand{\ea}{\end{eqnarray}}

\newcommand{\Bf}{{magnetic field}}
\newcommand{\Bfs}{{magnetic fields}}

\newcommand\eg{\textit{e.g.}}

\newcommand\lo{\mathrel{\raise.3ex\hbox{$<$}\mkern-14mu\lower0.6ex\hbox{$\sim$}}}
\newcommand\go{\mathrel{\raise.3ex\hbox{$>$}\mkern-14mu\lower0.6ex\hbox{$\sim$}}}

\begin{document}
\title{Cooling waves in pair plasma}

\author{Maxim Lyutikov,\\
Department of Physics, Purdue University, 
 525 Northwestern Avenue,
West Lafayette, IN
47907-2036   \\
and\\
Department of Physics and McGill Space Institute, McGill University, 3600 University Street, Montreal, Quebec H3A 2T8, Canada}

\begin{abstract}
We consider  structure and emission properties of a pair plasma fireball that cools due to radiation. At temperatures $T \geq 0.5 m_e c^2$ the cooling takes a form of clearly defined cooling wave, whereby the temperature and pair density experience a sharp drop within a narrow region. The surface temperature, corresponding to the location where  the optical depth to infinity reaches unity,  never falls much below $0.1 m_e c^2 \approx 50$ keV. The propagation velocity of the cooling wave is much smaller than the speed of light and decreases with increasing bulk temperature.
 \end{abstract}

 \maketitle
 
 \section{Introduction}
 
 Pair plasmas are important for various processes in astrophysics, like magnetars' burst and flares \citep[][]{TD95,2017arXiv170300068K} and  Gamma Ray Bursts  \citep{2004RvMP...76.1143P}. In laboratory, pair plasma can be created by powerful lasers \cite[][]{2006RvMP...78..591M}. After pair plasma is created it cools by expansion and radiative cooling. In this paper we consider radiative cooling of pair fireball at rest. 
 
As a prototype example,  consider a magnetar flare that releases $E_0\sim 10^{40}$ erg in a volume $r_0 ^3 \sim (10^6)^3$ cm$^3$ (these are typical values, Ref. \cite{2017arXiv170300068K}).  The fireball is confined by the magnetospheric \Bfs\ and can be assumed to have a constant radius.  Most of the energy is used to create pair plasma with temperature and density related by \cite{1982ApJ...253..842L,1982ApJ...258..335S} 
 \be
 n \lambda_C^3 =  \frac{\sqrt{2}}{\pi ^{3/2}}e^{-\frac{1}{\theta _T}} \theta _T^{3/2}
 \label{n}
 \ee
 where $\lambda_C=\hbar/(m_e c) $ is the electron Compton length, $\theta_T = T/(m_e c^2)$ and $T$ is temperature in energy units. 
  The corresponding temperature is given by 
   \be
   e^{-1/\theta_T} \theta_T^{3/2} \approx \left(\frac{E_0}{m_e c^2} \right) \left(\frac{\lambda_C}{r_0} \right)^3 \rightarrow \theta_T \approx  0.2
   \label{thetaT}
   \ee
  The temperature is not expected to exceed $m_e c^2$ by much, since pair plasma has large heat capacity - if more energy is added it is mostly used to create new pairs, not to increase thermal motion.
 The resulting optical depth
 \be
 \tau \sim \frac{E_0}{m_e c^2} \, \frac{ \sigma_T}{r_0^2} \approx 10^{10} \gg 1
 \ee
 Thus, overall the fireball is very optically  thick.

\begin{figure}[h!]
 \begin{center}
\includegraphics[width=.99\linewidth]{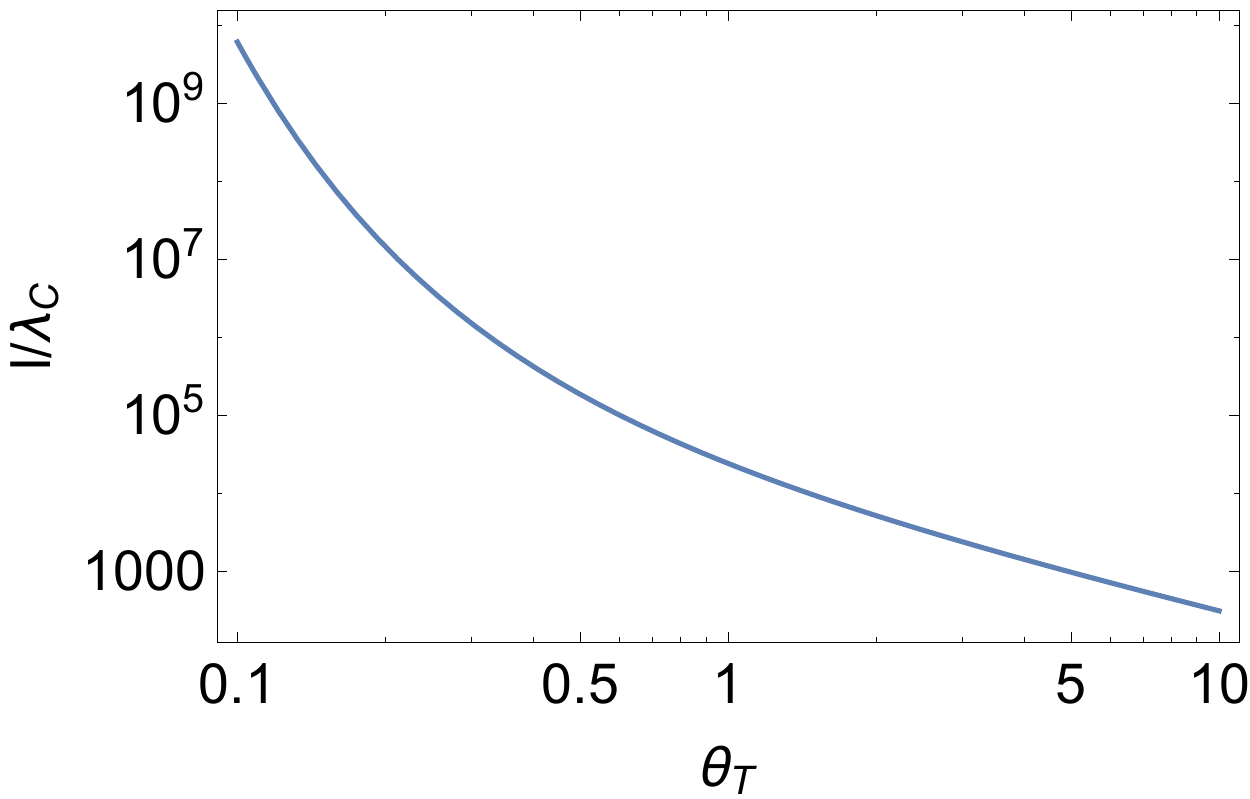}
\end{center}
\caption{Distance (in units of $\lambda_C$) that photons need to travel to achieve optical depth of unity in thermal pair plasma with temperature $\theta_T m_e c^2$. }
\label{ltau1}
\end{figure}

In other words,  in equilibrium pair plasma the distance $l_{\tau=1}$  that photons need to travel to achieve $\tau =1$ is, see Fig. \ref{ltau1},
\be
l_{\tau=1}= \frac{ 3 \sqrt{\pi}}{8 \sqrt{2} \alpha_f}\theta_T^{-3/2} e^{1/\theta_T}  \lambda_C
\ee

Fig. \ref{ltau1} indicates, that, qualitatively, as  the fireball cools the outer layers will become transparent at $\theta_T \approx 0.04 $ ($T= 20$ keV; for $r_0 =10^5-10^7$ cm).  More detailed calculations presented in this paper indicate somewhat higher surface temperature, \S \ref{Ts}.
In fact, thermodynamics equilibrium is violated in the outer parts, so that the density becomes nearly  constant, see \S \ref{non-eq}.

  The energy density at $\theta _ T \leq 1$ is dominated by the rest mass energy density. For example, the ratio of rest mass energy density to that of radiation is
  \be
  \frac{n m_e c^2}{4 \sigma_{SB} T^4/c} = \frac{15 \sqrt{2}}{\pi^{5/2}} \theta_T^{-5/2} e^{-1/\theta_T} \leq 1
  \ee
  where $\sigma_{SB}$ is the Stefan-Boltzmann constant. 
 For $\theta_T =1$ the above relation estimates to $0.44$. 
 
 After  pair fireball is created, it starts to cool. For an optically thick fireball the cooling will initially occur from a thin surface layer. As  temperature in the surface layer drops, it will absorb (actually, scatter in our case) hotter radiation from deeper layers. As a result, the temperature near the surface will start evolving with time.

 \section{Cooling of pair plasma}

 Time-dependent radiative transfer can occur in two somewhat  different regimes. Qualitatively, in the optically thick regime a local change of temperature (or more generally of the enthalpy) is related to the divergence of the radiation flux, which is in turn related to the gradient of the temperature (under assumption of local thermodynamic equilibrium, LTE). Thus, the corresponding equations involve one temporal derivative and two spacial derivatives. Due to the temperature dependence of the opacity, the resulting equation can, qualitatively, be either of the diffusion type, $\partial_t \propto \partial_x^2$, or of the Schrodinger equation,  $\partial_t \propto -  \partial_x^2$. These cases correspond, approximately, to two regimes of time-dependent radiative transfer: diffusive and that of a cooling wave. Typically, for a cooling wave-type behavior it is required that the opacity is a strongly 
 increasing function of temperature \citep{ZeldovichRaizer}.
 
 The theory of  cooling waves, in application to early stages of atmospheric nuclear explosions, was developed in Refs. \cite{Z1,Z2}, see also \cite{Bethe}.
In this paper we discuss applications of these ideas to the cooling of  stationary pair plasma. Stationarity, as opposed to  fast expansion, may be achieved by confining magnetic field in magnetars.

 Let us consider a regime of mildly relativistic plasma, $\theta_T \sim 1$. In this case the plasma energy density (enthalpy $w$) is dominated by the rest-mass energy density, $ w \approx m_e c^2 n$. Consider a half space filled with pair plasma. At time $t=0$ it is allowed to cool from the surface.  How will this cooling would proceed?
 
 As a first idealized problem we neglect any motion and assume that at any moment the plasma is in pair equilibrium, with density given by Eq. (\ref{n}).
 We are looking for the distribution of temperature $\theta _T(x,t)$.

Different layers of plasma exchange energy by radiation transfer; Thomson scattering being the dominant source of opacity. In the local diffusive approximation
the radiative flux at each point is
  \be
  F = - \frac{4}{3} \frac{a c T^3}{n \sigma_T} T' = -  \frac{\pi}{30}\frac{1}{\alpha_f^2}  \frac{1}{n \lambda_C^5}\theta_T^3 \theta_T' m_e c^3 =  \frac{\pi^{5/2}}{30 \alpha_f^2}  e^{1/\theta_T} \theta_T^{3/2}  \frac{ m_e c^3}{\lambda_C^2}\theta_T'
  \label{F}
\ee
where $a=4 \sigma_{SB}/c$;    the last step in (\ref{F})  assumes   that pair density is in equilibrium; $\alpha_f$ is the fine structure constant. Prime denotes differentiation with respect to $x$. 
Eq. (\ref{F}) assumes diffusive propagation of light - it is justified in the optically thick region.

 Energy continuity  requires
\be
\dot{ w} +F'=0
\label{energy}
\ee
where $w$ is the enthalpy, dot denotes time derivative.
For $w=  m_e c^2 n$ and using the expression for the flux (\ref{F}),  Eq. (\ref{energy}) takes the form
\be
\dot{\theta}_T= \kappa \frac{e^{2/ \theta_T}}{2+3 \theta_T} 
\left(2 \theta_T^2 \theta_T^{\prime \prime} -(2 -3 \theta_T) (\theta_T^\prime)^2 \right)
\label{main}
\ee
This is the main equation that describes the cooling of pair plasma due to escaping radiation. 

Equation (\ref{main}) is a non-linear diffusion/Schrodinger type equation (first order  derivative in time and second order spacial derivative) for $\theta_T(x,t)$. It is not clear at first what would be its asymptotic behavior - like a diffusive wave (if the right hand side resembles more the diffusion equation) or a cooling propagating wave,
 (if the right hand side resembles more the Schrodinger equation). To study it's behavior we employ two self-similar parametrization: (i) diffusive, when all quantities depend on $z= x/\sqrt{t}$ and (ii) that of a cooling wave (CW), all quantities depend on $z= x- \beta_{CW} c t$. Getting a bit ahead, it is the CW approach that gives physically meaningful  results.
 
 \subsection{Diffusive propagation}
Assuming that $\theta_T\equiv \theta_T(z=x/\sqrt{t})$ we find that the main equation (\ref{main}) allows a self-similar parametrization: 
\be
\frac{\theta_T^{\prime \prime}}{\theta_T^{\prime}}=\frac{z e^{-\frac{2}{\theta _T}} \left(3 \theta _T+2\right)}{4 \kappa  \theta _T^2}+
\frac{2-3 \theta _T}{2 \theta _T^2} \theta_T^{\prime}
\label{diff}
\ee

It is useful at this point to compare Eq. (\ref{diff})  with the simple diffusion equation $\dot{y}= y^{\prime \prime}$. Parametrization $y (x/\sqrt{t})$ gives
$y{\prime \prime}/y^{\prime}= - z/2$, with a solution   $y \propto {\rm Erf} (z)$. A related Schrodinger-type equation, $y{\prime \prime}/y^{\prime}=  z/2$  (different sign of the rhs) has a solution exponentially divergent for $z \rightarrow \infty$. It is not clear at first sight what type of equation  (\ref{diff})   is.

\begin{figure}[h!]
 \begin{center}
\includegraphics[width=.99\linewidth]{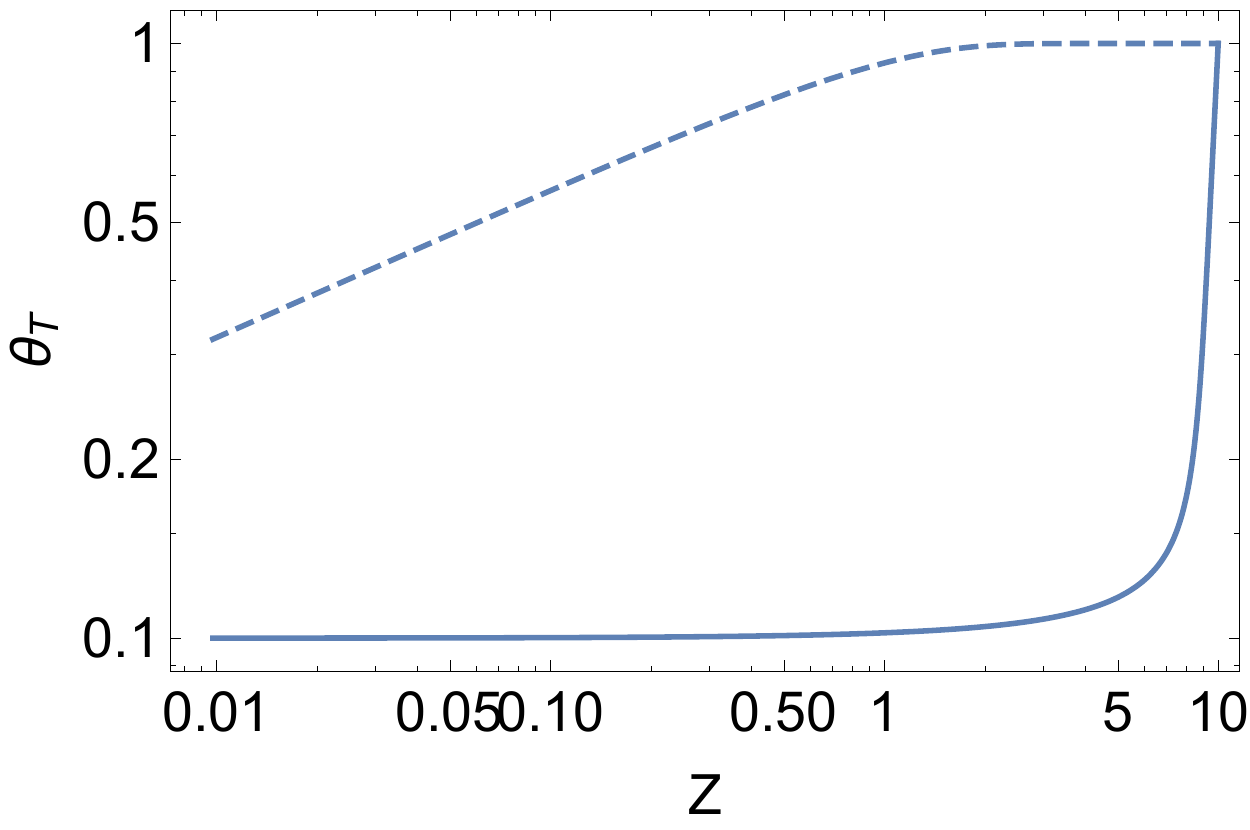}
\end{center}
\caption{Self-similar solutions for $\theta_T(x/\sqrt{t})$ for pair plasma (solid line) and constant density high temperature plasma (dashed line) with boundary conditions
$\theta_T(10)=1$ and  $\theta_T(0)=0.1$ and $\kappa=\kappa_1=1$.
In case of the pair plasma the  solution is exponentially divergent at the outer boundary (and thus is unphysical), while for constant density the solution resembles a diffusive wave.
}
\label{self-similar-compare}
\end{figure}

Numerical solutions of Eq. (\ref{diff}) (\eg, with fixed value of $ \theta _T$ at $z=0$ and  large $z \rightarrow  \infty$) show exponential divergence, qualitatively similar to self-similar solution of Schrodinger-type equation, Fig. \ref{self-similar-compare}, solid line.

To highlight the point that very similar set-ups can show qualitatively different behavior (that of diffusive cooling and that of the cooling wave), let us compare the case of pair plasma with the case of just high temperature plasma with constant density. In both cases radiation transfer is assumed to be dominated by Thomson scattering. 
For normal plasma $w= m_e c^2 n T$ the energy continuity  equation (\ref{energy}) takes the form
\be
\dot{\theta}_T = \frac{1}{2 \kappa_1} \left(3 (\theta_T')^2 + \theta_T \theta_T^{\prime \prime} \right) \theta_T
\ee
where $\kappa_1= \pi/( 15 \alpha_f ^3 c m_e \lambda_C^5 n^2)$. The self-similar anzats gives
\be
\frac{\theta_T^{\prime \prime}}{\theta_T^{\prime}}= -\frac{1}{ \kappa_1} \frac{z}{\theta_T^{3}} - 3 \frac{\theta_T^{\prime}}{\theta_T^{2}}
\label{diff1}
\ee
The overall form of Eq. (\ref{diff1}) is very similar to  (\ref{diff}), yet the solutions are qualitatively different,   Fig. \ref{self-similar-compare}, dashed line. In this case the cooling takes a form of a diffusive spreading.

  Thus we conclude that that cooling of the pair plasma is not a diffusive self-similar process, but proceeds in a form of a propagating cooling wave.

As we demonstrate below, CW approximation breaks down at temperatures $\theta_T \leq 0.5$. In that regime the propagation is likely to be diffusive, but  not self-similar.  Qualitatively,   in the regime   $\theta \leq 1$, after time $t$ the cooling affects distance up to
   \be
   x \sim e^{1/\theta_T}  \sqrt{  \theta _T  \kappa t }
   \ee
There is sensitive, exponential,  dependence on $\theta_T$.

 \section{Propagation of cooling wave in pair plasma.}
 \subsection{Structure of the cooling wave}
 Opacity of pair plasma to Thomson scattering increases with temperature, as more and more pairs are produced at high $T$. Qualitatively, this increase of opacity with temperature is similar to the case of nuclear explosion in  air \cite{ZeldovichRaizer}. It is well know that in that case cooling takes a form of a cooling wave, whereby the temperature evolves smoothly in the optically thick part and drops suddenly at the transition $\tau \approx 1$. 
 
 A cooling wave is launched into the bulk of the plasma. At this point we are interested in the asymptotic dynamics of that cooling wave.
  Following \cite{ZeldovichRaizer} we assume that all quantities (temperature)  depend on $x - \beta_{CW} c t$ where  $\beta_{CW}$ is the speed of the cooling wave,
 $T(x - \beta_{CW} c t)$.

Using  (\ref{F}), the energy equation (\ref{energy})  can be integrated
\be
 F= \beta_{CW}  (w_0-w)
 \label{F1}
 \ee
 where $ w_0= me c^2 n_0 $ is the value of  enthalpy far ahead of the CW, where radiative flux is zero.

\begin{figure}[h!]
 \begin{center}
\includegraphics[width=.99\linewidth]{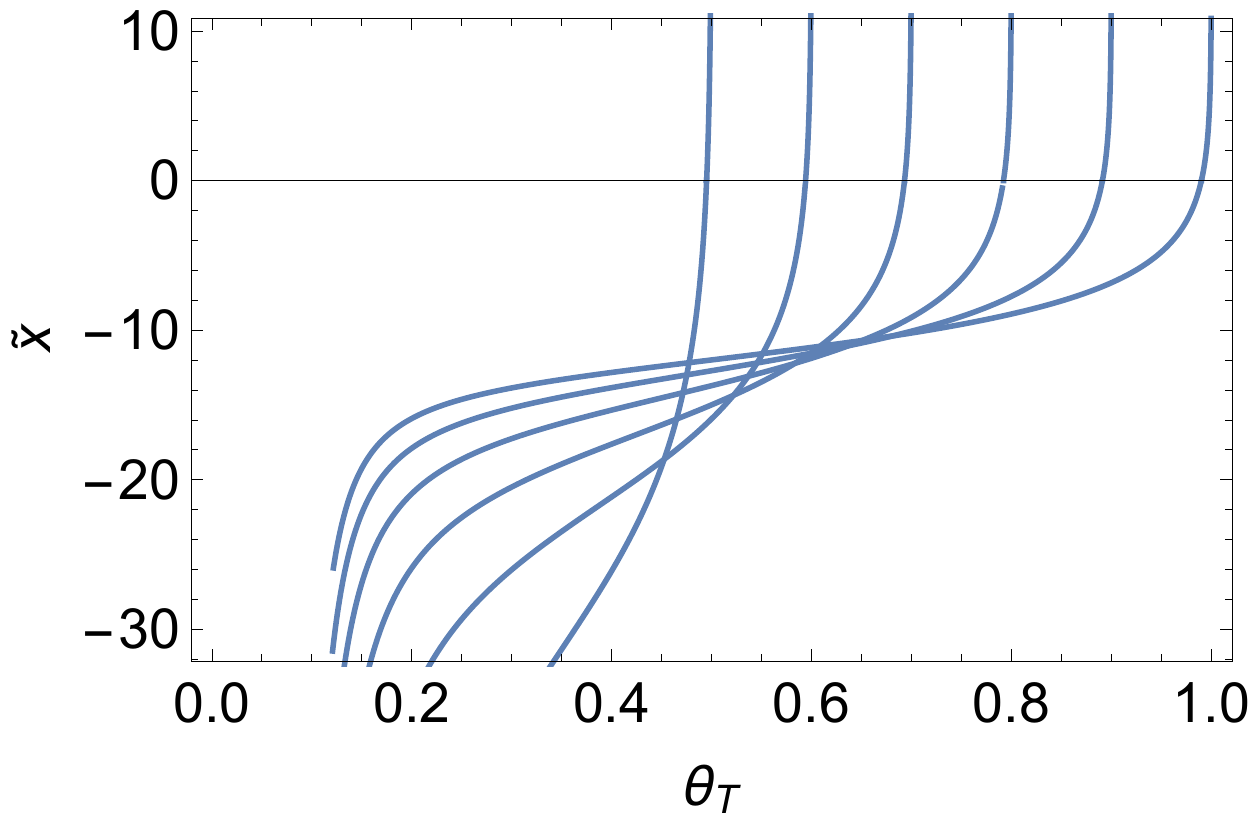}
\end{center}
\caption{Structure of the cooling wave in pair plasma. Plotted is the normalized coordinate $\tilde{x} = ( 60 \alpha_f^2/\pi^4 ) (x/\lambda_C) $  (and assumed $\beta_{CW}=1$ for the plot) as a function of temperature $\theta_T$
for various $\theta _{T_0}=0.5,\, 0.6 ,\, 0.7,\, 0.8 ,\, 0.9,\, 1$. For  $\theta _{T_0} \geq 0.5$ there is a clearly defined sharp transition region, where the temperature drops -  this is the cooling wave. For smaller $\theta _{T_0}$ the CW becomes very broad, losing the whole notion of a well-localized structure. 
The wave propagates to large positive $x$, where $\theta_T \rightarrow \theta _{T_0}$. Location of $\tilde{x} =0$ is chosen to be where temperature is $0.99\theta _{T_0}$.  (Minor breaks near $\tilde{x} =0$ are due to different procedures used to plot: numerical integration for $\tilde{x} <0$ and analytical for $\tilde{x} >0$, see Eq.  (\protect\ref{F3}). }
\label{xofthetaT}
\end{figure}

Using (\ref{F}), Eq. (\ref{F1}) takes the form
\ba &&
\theta_T'= - {\beta_{CW}} {\cal F}( \theta _T,\theta _{T_0})
\nn && 
\tilde{x} =\frac{\pi^4 }{60 \alpha_f^2} \frac{ x}{\lambda_C}
\nn &&
{\cal F}( \theta _T,\theta _{T_0})= -  e^{-\frac{1}{\theta _{T_0}}-\frac{2}{\theta _T}} \left(e^{\frac{1}{\theta _{T_0}}}
   \theta _T^{3/2}-e^{\frac{1}{\theta _T}} \theta _{T_0}^{3/2}\right)
   \label{F2}
   \ea
   where derivative now is with respect to $\tilde{x} = ( 60 \alpha_f^2/\pi^4 ) (x/\lambda_C) $.
   For a given $\beta_{CW} $ equation (\ref{F2}) can be integrated in quadratures to find $x(\theta_T)$, Fig. \ref{xofthetaT}.
For sufficiently high bulk temperatures,  $\theta_{T_0} \geq 0.5$, there is a clear sharp increase of a temperature occurring in a narrow spacial range - this is the cooling wave. On the other hand, for smaller $\theta_{T_0} \leq 0.5$, the temperature change occurs in a broad region, reminiscent more of a diffusive relaxation of the temperature. 
      
In the limit $\theta \rightarrow \theta _{T_0}$ we find  $  {\cal F} \propto  \frac{\epsilon  e^{-\frac{2}{\theta _{T_0}}} \left(3 \theta _{T_0}+2\right)}{2
   \sqrt{\theta _{T_0}}} \Delta \theta_T$, where $\Delta \theta_T= \theta _{T_0}- \theta _{T}$. Thus,
   \be
   x \propto   \ln1/ \Delta \theta_T
   \label{F3}
   \ee

   \subsection{Surface temperature}
   \label{Ts}
   
\begin{figure}[h!]
 \begin{center}
\includegraphics[width=.49\linewidth]{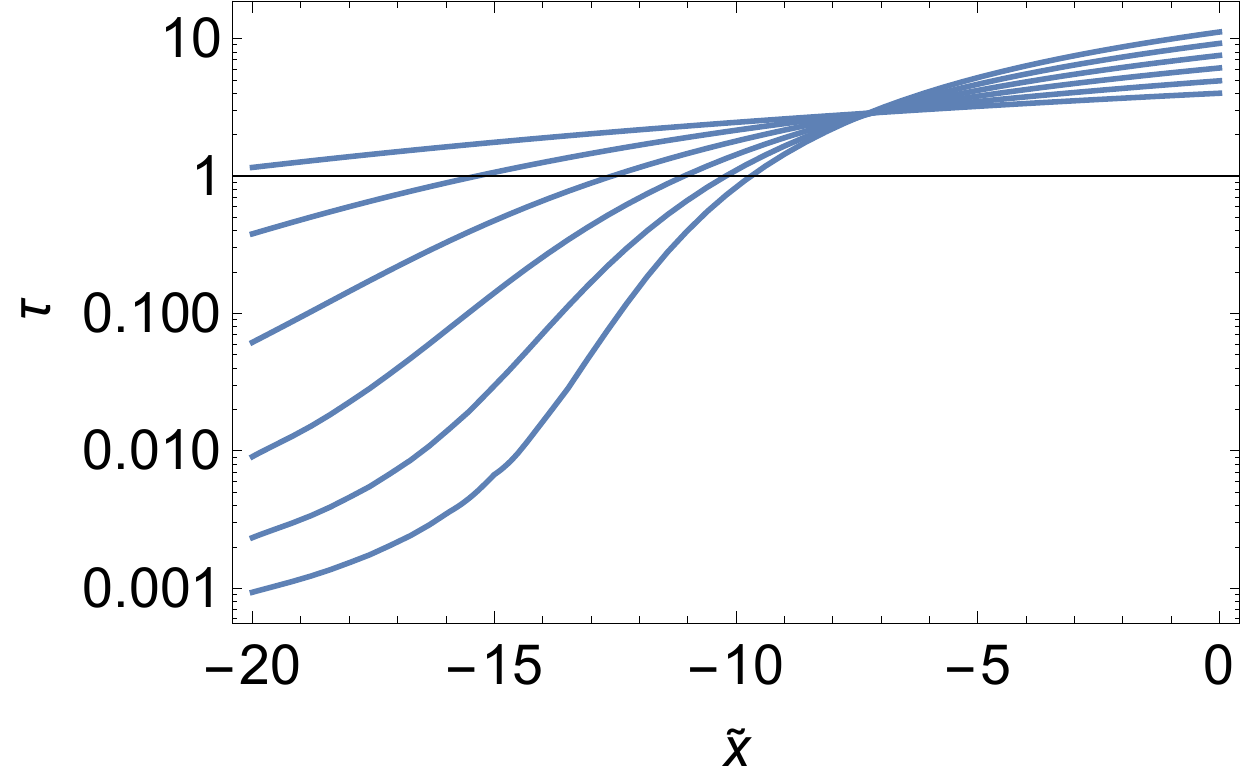}
\includegraphics[width=.49\linewidth]{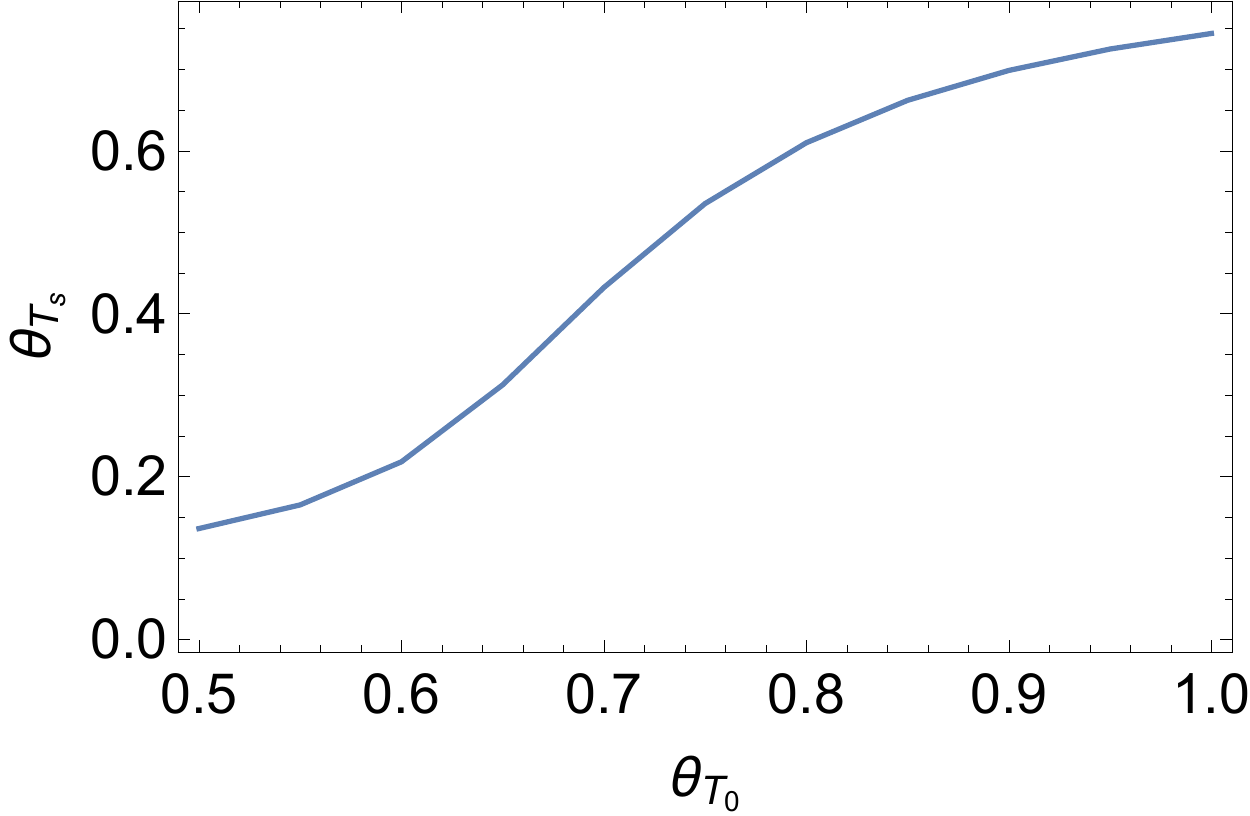}
\end{center}
\caption{Left Panel: optical depth to a given $\tilde{x}$ for different $\theta_{T_0}=0.5,\, 0.6...1$ (bottom to top). Right Panel: surface temperature  $\theta_{T_s}$ as a function of  $\theta_{T_0}$. Importantly, $\theta_{T_s}$ never goes below $\sim 0.2$. }
\label{TsofT0}
\end{figure}

   The observed surface temperature is determined by the condition of optical depth $\sim 1$. As a major simplification, that it likely to be violated in applications, let us assume that even in the optically thin regime the density of pairs is given by the thermodynamic equilibrium with radiation (see \S \ref{non-eq} for discussion of non-equilibrium effects).  Then the condition $\tau =1$ is given by
  \be
  1 = \int \sigma_T n dx =  \frac{2 \sqrt{2} \pi^{7/2}}{ 45 \beta_{CW}} \int e^{1/\theta_T} \theta_T^{3/2} d \tilde {x}
  \ee
    From the previous, Fig. \ref{xofthetaT}, we know the distribution of temperatures and, under LTE assumption, a distribution of densities. We can then calculate an optical depth to a given point $\tilde {x}$ and the corresponding temperature $\theta_{T_s}$, Fig. \ref{TsofT0}.  
 Importantly, $\theta_{T_s}$ never goes below $\sim 0.15$.  This is due to the exponential dependence of pair density on the temperature - cold plasma provides virtually no contribution to the optical depth.

  \subsection{Propagation velocity of cooling wave}
  \label{betaCW1}
  
\begin{figure}[h!]
 \begin{center}
\includegraphics[width=.99\linewidth]{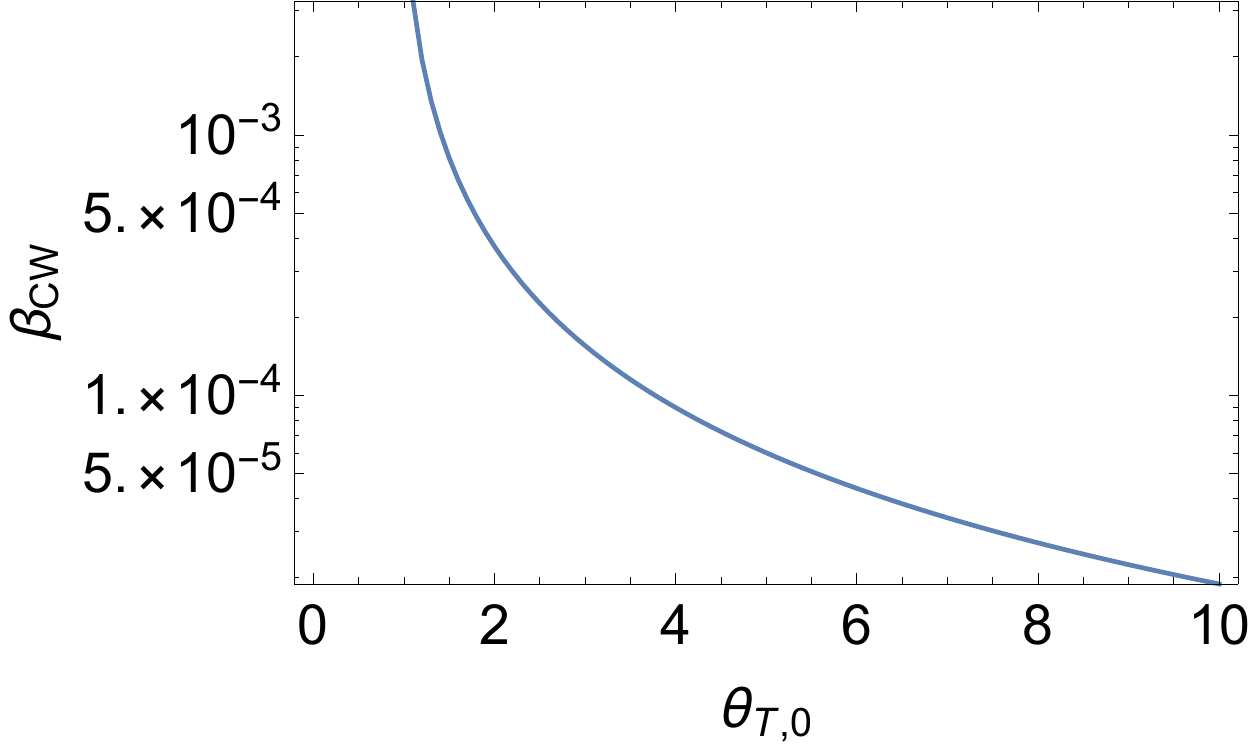}
\end{center}
\caption{Speed of the cooling wave as a function of the basic temperature $\theta_{T,0}$. For $\theta_{T,0}\leq 1$ the employed scheme is unstable, presumably due to the fact that at those temperature a notion of a cooling wave is less clearly define (temperature transition occurs over a broad range, see Fig. \protect\ref {xofthetaT}).}
\label{betaCW}
\end{figure}

  Above in \S \ref{Ts} we have determined the location and the surface temperature. Now we are in a position to calculate the propagation velocity of cooling wave $\beta_{CW}$.  At the surface the radiation flux is 
  \be
  F=  \frac{4 \sigma_{SB}}{c} T_s^4 = \frac{\pi}{15} \frac{m_e c^2}{\lambda_C^3} \theta_{T,s}
  \label{F5}
  \ee
This should equal the radiation flux given by Eq. (\ref{F1}). Eq. (\ref{F1}) depends on the velocity of the cooling wave explicitly, 
while Eq. (\ref{F5}) depends on the velocity of the cooling wave implicitly, via the surface temperatures $\theta_{T,s}$, Eq. (\ref{main}).
The results of the relaxation calculations are pictured in Fig. \ref{betaCW}.

Importantly, the velocity of the cooling wave, Fig. \ref{betaCW}, turns out to be  much smaller than what could be expected from a simple estimate,
\be
\beta _{CW,0} = \frac{\sigma_{SB} T^4}{m_e c^3 n} = \frac{\pi^{5/2}}{60 \sqrt{2}} e^{1/\theta_{T,s}}\theta_{T,s}^{5/2}
\ee
For example, at $\theta_T =1$ we find $\beta_{CW} = 10^{-1}$.

\section{Non-equilibrium effects -  pair freeze-out}
\label{non-eq}
Annihilation rate in mildly relativistic pair plasma is \citep{1982ApJ...258..335S}
\be
\dot{n} = -  \pi n^2 c r_e^2.
\ee
In an optically thin region the density will then evolve according to 
\be
n= \frac{n_i}{1+  \pi n_i c r_e^2  t} \rightarrow \frac{1}{\pi  c r_e^2  t}
 \label{nn}
 \ee
 where $n_i$ is the density at the moment of plasma becoming optically thin.
The density thus will decrease very slowly - this is a pair freeze-out. There is expectation of 511 keV emission from the plasma, with total energy $r_0^3 n_{i} m_e c^2$, with fast decreasing  luminosity $\propto 1/t^2$, Eq.  (\ref{nn}).

 The optical depth through the  freeze-out regions, assuming that post-freeze out emissivity is zero, and cooling wave speed is $\beta_{CW} c$, can be estimated as
 \be
 \tau_f = n(t) \sigma_T \beta_{CW} c  t =   \frac{8}{3 } \beta_{CW}
 \ee
 where we used (\ref{nn}) to estimate the density.
 Thus, for slow propagating cooling wave, see \S \ref{betaCW1},  in pair plasma, $\beta_{CW} \ll 1$, the resulting cooled plasma is optically thin,  $\tau_f \ll 1$. Recall, Fig. \ref{betaCW}, that the velocity of the cooling wave decreases with increasing internal temperature. Thus, the higher $\theta_{T,0}$ is, the less optically thick the envelope is. 
 On the other hand, for $\theta_{T,0} \leq 1$ the non-LTE envelop might strongly affect the appearance of the pair  fireball.

\begin{figure}[h!]
 \begin{center}
\includegraphics[width=.99\linewidth]{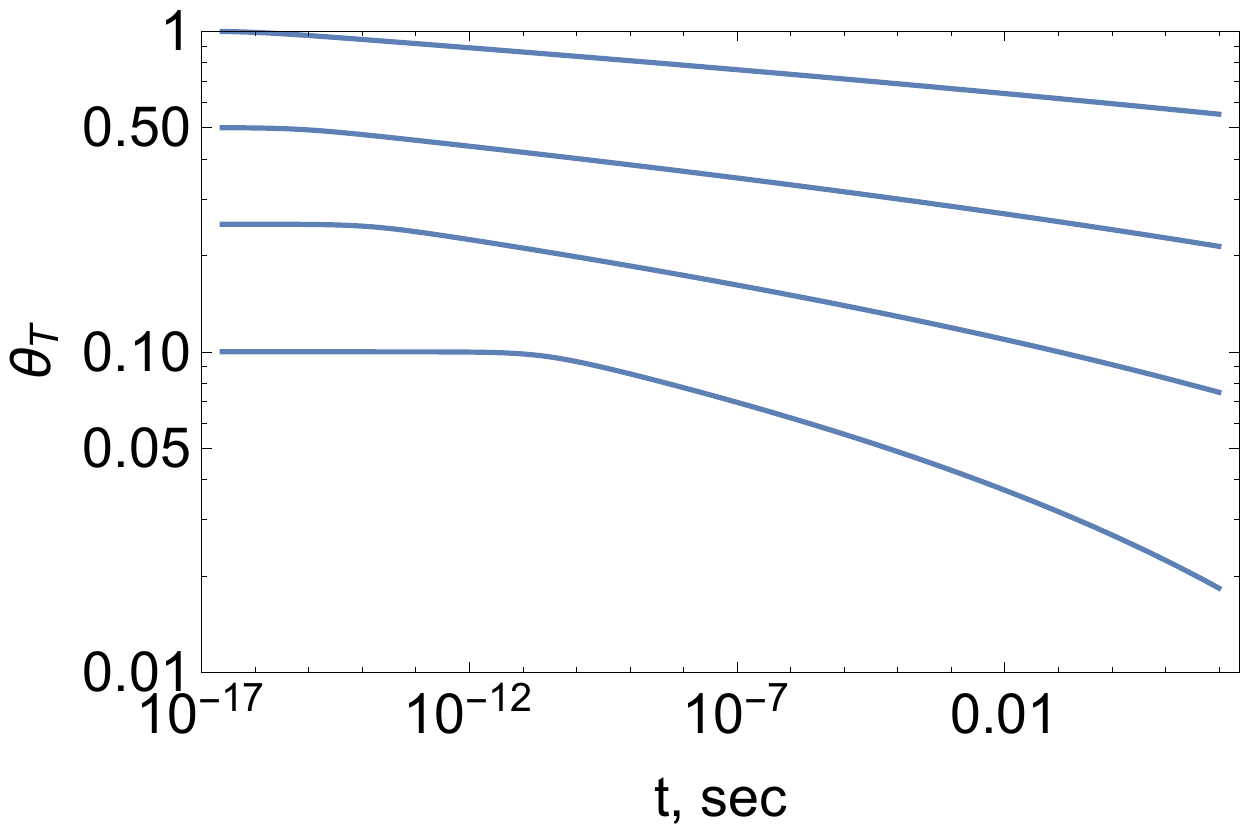}
\end{center}
\caption{Evolution of the temperature in the optically thin regime for different $\theta_{T_0} =0.1,
\, 0.25,\, 0.5,\, 1$ (bottom to top).}
\label{Toft}
\end{figure}

After becoming optically thin the plasma will cool via free-free emission, losing energy at a rate 
\be
J_{tot} = \frac{16}{3} \frac{ e^6 \sqrt{T} n^2 g_{ff}}{\hbar c^3 m_e^{3/2}}
\ee
where $g_{ff} $ is the Gaunt factor.  Thus, plasma cools logarithmically  slow,
\be
\theta_T= \theta_{T_s} \left(1 - \frac{8 g_{ff} \alpha_f}{3 \pi \sqrt{\theta_{T_s}}} \ln \left( 1+ \sqrt{\frac{2}{\pi}} e^{-1/\theta_{T_s}} \theta_{T_s}^{3/2} \frac{ c \alpha_f^2 t}{\lambda_C}  \right) \right)^2 ,
\ee
(for $\theta_{T_s} \leq 1$ the factor under the logarithm is exponentially large), Fig. \ref{Toft}.
Qualitatively, on time scale $\alpha_f^2 \lambda_C/c$ the temperature remains constant, and then decreases logarithmically. 

   \section{Discussion}
   In this paper we considered the radiative cooling of a pair plasma fireball. We found that for temperatures $\theta_T \geq 0.5$ the cooling takes a form of a clearly defined cooling wave, propagating with sub-relativistic velocities. These effects are likely to be important for magnetar flares. We find that the temperature of the  surface of the cooling wave never falls below $0.15 m_e c^2$. Pair freeze-out in the outer parts of the CW may create a slowly cooling shroud that can reach $\sim 0.03 m_e c^2$. At smaller temperatures there is just no enough pairs to cool the plasma
   
   Observationally, magnetar flares can reach surface temperatures as low as 3 keV $= 0.006 m_e c^2$ \cite{2017arXiv170300068K}. This is difficult to achieve in the given model.  Another important step is the effects of \Bf.   We plan to address the effects of nearly critical \Bf\ in a subsequent publication.  
    
 I would like to thank Andrew Cumming and Christopher Thompson for discussions. This work had been supported by   NSF  grant AST-1306672 and DoE grant DE-SC0016369.

 \bibliographystyle{apsrev}
  \bibliography{/Users/maxim/Home/Research/BibTex}

\end{document}